\RequirePackage{fix-cm}
\documentclass[smallextended]{svjour3}       % onecolumn (second format)
\smartqed  % flush right qed marks, e.g. at end of proof
\usepackage{graphicx}
\usepackage{natbib}
\usepackage{placeins}
\usepackage{epsf}
\usepackage[flushleft]{threeparttable}
\usepackage{amsmath,amssymb}      %for \lesssim and \gtrsim symbols
\newcommand{\be}{\begin{equation}}
\newcommand{\ee}{\end{equation}}

\def\***#1{\textbf{\textsf{*** #1 ***}}}

\journalname{SSRv}
\begin{document}

\title{X-ray spectroscopy of galaxy clusters: beyond the CIE modeling %\thanks{Grants or other notes
}

%\titlerunning{X-ray spectroscopy}        % if too long for running head

\author{Liyi Gu         \and
        Irina Zhuravleva \and
        Eugene Churazov \and  
        Frits Paerels   \and 
        Jelle Kaastra \and  
        Hiroya Yamaguchi %etc.
}


\institute{L. Gu \at
              RIKEN High Energy Astrophysics Laboratory, 2-1 Hirosawa, Wako, Saitama 351-0198, Japan \\
              SRON Netherlands Institute for Space Research, Sorbonnelaan 2, 3584 CA Utrecht, the Netherlands \\
              \email{liyi.gu@riken.jp}           \\
           \and
           I. Zhuravleva \at
              Kavli Institute for Particle Astrophysics and Cosmology, Stanford University, 452 Lomita Mall, 
              Stanford, California 94305-4085, USA \\
              Department of Physics, Stanford University, 382 Via Pueblo Mall, Stanford, California 94305-4060, USA \\
           \and 
           E. Churazov \at
           Max-Planck-Institut f$\rm \ddot{u}$r Astrophysik, Karl-Schwarzschild-Strasse 1, 85741 Garching, Germany \\
           Space Research Institute (IKI), Profsoyuznaya 84/32, Moscow 117997, Russia \\
           \and 
           F. Paerels \at
           Columbia Astrophysics Laboratory and Department of Astronomy, Columbia University, 538 W. 120th St., New York, NY 10027, 			USA \\
           \and 
           J. Kaastra \at 
           SRON Netherlands Institute for Space Research, Sorbonnelaan 2, 3584 CA Utrecht, the Netherlands \\
           Leiden Observatory, Leiden University, Niels Bohrweg 2, 2300 RA Leiden, the Netherlands \\
           \and 
           H. Yamaguchi \at 
           Institute of Space and Astronautical Science (ISAS), Japan Aerospace Exploration Agency (JAXA), Kanagawa 252-5210, Japan \\       }

\date{Received: date / Accepted: date}

\maketitle

\begin{abstract}

X-ray spectra of galaxy clusters are dominated by the thermal emission from the hot intracluster medium. In some cases, besides the thermal component, spectral models require additional components associated, e.g., with resonant scattering and charge exchange. The latter produces mostly underluminous fine spectral features. Detection of the extra components 
therefore requires high spectral resolution. The upcoming X-ray missions will provide such high resolution, and will
allow spectroscopic diagnostics of clusters beyond the current simple thermal modeling. A representative science case is resonant
scattering, which produces spectral distortions of the emission lines from the dominant thermal component. 
Accounting for the resonant scattering is essential for accurate abundance and gas motion 
measurements of the ICM. The high resolution spectroscopy might also reveal/corroborate a number of 
new spectral components, including the 
excitation by non-thermal electrons, the deviation from ionization equilibrium, and charge exchange 
from surface of cold gas clouds in clusters. Apart from detecting new features, future high resolution spectroscopy
will also enable a much better measurement of the thermal component. Accurate atomic database and appropriate
modeling of the thermal spectrum are therefore needed for interpreting the data.

\end{abstract}

\section{Introduction}
\label{intro}
%Some introduction about cluster science with
%high resolution X-ray spectrum.

Galaxy clusters are the largest bound systems in the Universe. The dominant baryonic matter in clusters is in the form
of the intracluster medium (hereafter ICM), a tenuous ($10^{-5} - 10^{-2}$ cm$^{-3}$), hot (10$^{7-8}$ Kelvin), and large scale ($\sim$Mpc) 
plasma in quasi-hydrostatic equilibrium within the dark matter potential. ICM reveals itself by emitting mostly thermal
X-rays, which has been extensively studied for assessing the physical properties of the ICM, and for tracing the cosmological
structure formation in the Universe.

Many achievements in the studies of ICM were made possible by the advent of X-ray spectroscopic instruments. 
The line emission of highly-ionized iron was first discovered with the Ariel V spectrum (spectral resolution $R\sim 6$) of 
the Perseus cluster, establishing the thermal origin of the ICM \citep{1976MNRAS.175P..29M}. The improved spectral resolution with
CCD-type spectrometers ($R=10-60$) led to the discovery of other fundamental elements (O, Ne, Mg, Si, 
S, Ar, Ca, and Ni) in the ICM \citep{1996ApJ...466..686M}. The reflection grating spectrometer ($R=50-100$ for spatially extended sources)
onboard {\it XMM-Newton} made a breakthrough in the soft X-ray spectroscopy. One of its major discoveries is the lack of
massive accumulation of the cold gas in cool core clusters \citep{2001A&A...365L.104P}. More recently, X-ray micro-calorimeters have been developed. 
The micro-calorimeter onboard {\it Hitomi} ($R\sim 1250$) measured directly velocities of ICM random motions and detected the Fe-peak elements Cr, Mn, and Ni for
the first time \citep{2016Natur.535..117H, 2017Natur.551..478H}.

ICM is far from being quiescent. High-resolution imaging of cluster cores revealed
the presence of optical-line emitting filaments, cavities, and weak shocks possibly created by the AGN feedback; on larger scales, radio-luminous shocks by mergers have been observed. The implied energetic
thermodynamical processes are expected to strongly affect the cluster evolution.
Some of the non-thermal processes could, in theory, be detected by X-ray spectrometers, as they would produce 
characteristic features in the spectrum. In practice, however, firm detections with present 
X-ray instruments are still limited, as the X-ray signatures are often faint and tend to be swamped 
by the dominant ICM thermal signal.

Future high-resolution X-ray spectrometers, e.g., the X-ray imaging and spectroscopy mission (XRISM, launch 2021) and Athena (launch 2030), will explore the ICM physics beyond the current standard 
thermal view. In this review we summarize effects that introduce additional features in the spectra of galaxy clusters and can be observed with future X-ray spectrometers.
In particular, the resonant scattering effect (\S\ref{sec:1}) will be possible to detect in
many clusters, and will be widely used as a proxy of random motion in the ICM. In addition, the 
presence of non-thermal (\S\ref{sec:2}), non-equilibrium ionization (\S\ref{sec:3}), and charge exchange 
(\S\ref{sec:4}) phenomena could be
addressed through measurements of particular line emission. 

Apart from detecting new X-ray features, it is also important to model properly the known thermal emission. As more high-resolution X-ray spectra become available, more accurate atomic data and plasma models will be urgently needed. Here, we also review the status of current
atomic codes (\S\ref{sec:5}), the achievements and challenges in the work with the 
{\it Hitomi} spectrum of the Perseus cluster.

\section{Resonance scattering}
\label{sec:1}

%$\rightarrow$ Zhuravleva, Paerels, Churazov

%1.1 overview of RS physics/history (1-2 pages) \\
%-- quick summary of the 2010 review, focus on the basics, point discoveries and implications to the gas motion review \\
%1.2 improved numerical modeling of the RS spectrum in cluster core (~2 pages) \\
%-- basic approach \\
%-- full lists of the optically-thick lines for galaxy-level and cluster-level spectra \\
%-- compare the RS measurements among different approaches (full simulation of radiative transfer, ionized absorption) and address the uncertainties  \\
%1.3 caveat/uncertainties of the atomic physics (1-2 pages) \\
%-- accuracy of theoretical calculation and laboratory measurements of the Fe line emissivities \\
%-- systematics due to other atomic processes (charge exchange...) \\

Resonant scattering (RS) is a process of absorption and re-emission of a photon, having the energy close to a resonant transition of an atom or ion.  The resonant scattering is especially interesting in the context of galaxy clusters because it can affect measured element abundances in the cluster cores and, also, serve as an independent proxy for gas motions \citep{1987SvAL...13....3G, 2010SSRv..157..193C}. Measuring line width using high resolution X-ray spectra is the most direct way of probing the motions, but for extended sources such data are still sparse \citep{2016Natur.535..117H}.
Moreover, given the relatively low energy resolution of future X-ray calorimeters at the Fe-L energies 
\citep{2016SPIE.9905E..0UT}, resonance scattering will
remain an important tool for measuring gas motion in elliptical galaxies and galaxy groups, which emit primarily in the soft X-ray band.

%RS offers one of the complementary routes to probe the velocities and also enlarges the toolkit for studying the ICM \citep[see][for review]{2010SSRv..157..193C}.

\subsection{Optical depth and line width}
Despite their large size ($L\sim$Mpc), galaxy clusters contain very low density gas ($n_e\sim n_p\sim 10^{-5}-10^{-2}\; {\rm cm^{-3}}$). This results in a small Thomson optical depth  ($\tau_T\lesssim 10^{-3}$) and essentially zero free-free opacity in X-rays. However, the optical depth near the resonant transition of astrophysically abundant elements can be of the order of unity \citep{1987SvAL...13....3G}. The optical depth at the center of a line can be written as
\be
\tau_{RS}=n_p L Z
\,\delta_i(T) \frac{\sqrt{\pi}hr_e cf}{\Delta E_D},
\label{eq:tau}
\ee where $L$ is the characteristic size, $Z=\frac{n_Z}{n_H}$ is the abundance of an element with respect to hydrogen; $\delta_i(T)$ is the fraction of a given ion, set by the ionization balance; $h$, $r_e$, $c$ are the Planck constant, classical electron radius, and the speed of light, respectively; $f$ is the absorption oscillator strength of a given atomic transition. $\Delta E_D$ is the Doppler width of the line, set by the ion thermal velocities and by the ``turbulent'' velocities within the observed region
\be
\Delta E_D=E_0\left[\frac{2kT_e}{Am_p c^2}+\frac{2V_{1,turb}^2}{c^2}\right]^{1/2},
\label{eq:de}
\ee where $A$ is the atomic weight of the corresponding element, $m_p$ is the proton mass and $V_{1,turb}$ is the RMS of gas velocities in one direction. A natural width of the line could also be included in the expression for $\Delta E_D$, but for the ICM conditions it is usually sub-dominant.

Three things are immediately clear from eqs.~(\ref{eq:tau}) and (\ref{eq:de}). Firstly, it is likely that due to the factor $Z \,\delta_i(T)$ in eq.~(\ref{eq:tau}), the optical depth will be the largest for [He-] and [Ne-] like ions of the most abundant elements.

 Secondly, in the expression for the line width, the ion thermal broadening is ``attenuated'' by the atomic weight $A$, which makes $\Delta E_D$ and, therefore, the optical depth $\tau_{RS}$, sensitive to the gas motions, even if they are subsonic. Thirdly, the brightest lines in the optically thin plasma are often associated with resonant transitions. Therefore, the very same line photons will be the most strongly affected by the resonant scattering.

\subsection{Phase function and polarization}
\label{sec:phase}
For any transition, the resonant scattering can be represented as a
combination of two processes: isotropic scattering with a weight $w_1$
and Rayleigh scattering with a weight $w_2=1-w_1$ \citep{1947ApJ...106..457H}.
The weights $w_1$ and $w_2$ depend on the total angular
momentum $j$ of the ground level and on the difference between
total angular momenta of the excited and ground levels $\Delta j$
(=$\pm 1$ or $0$).  If the initial radiation is unpolarized, the probability of the photon scattering by angle $\theta$ is
\begin{equation}
P(\mu)=\frac{1}{4\pi}\left[w_1+\frac{3}{4}(1+\mu^2)w_2\right],
\label{eq:phase}
\end{equation}
where $\mu=\cos \theta$. The isotropic component of the phase function does not induce any polarization, while the Rayleigh component does.  If polarization is of interest, then a full scattering matrix for Stokes parameters has to be used \citep[][]{1947ApJ...106..457H,Chan50}. Given that more line photons are  generated in the dense cluster cores,  the polarization arises\footnote{for transitions with non-zero Rayleigh component} in the cluster periphery, with the polarization plane orthogonal to the radial direction \citep{2002MNRAS.333..191S}. The resonant line of the He-like triplet $w_1=0$ and $w_2=1$, and, therefore, is the most promising target for future polarimeters.
The magnitude of the polarization is sensitive to the line broadening via the optical depth, and, therefore, to the turbulent velocities \citep{2010MNRAS.403..129Z}. In many practical applications, one can ignore polarization effects and use an isotropic phase function for all transitions.  The errors introduced by this approximation are often smaller than the uncertainties in the model, e.g., an assumption of a spherically symmetric cluster.

\subsection{Surface brightness and line profile}
Apart from the polarization, the resonant scattering leads to (i) a modification of the surface brightness profile of a cluster in the resonant line and (ii) changes in the line spectrum.

\begin{figure}
\centering
\begin{minipage}{0.49\textwidth}
\includegraphics[trim= 1mm 5cm 20mm 2cm,
  width=1\textwidth,clip=t,angle=0.,scale=0.86]{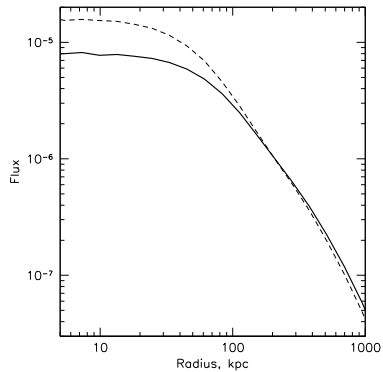}
\end{minipage}
\caption{Radial profiles of the He-like iron $K_\alpha$
  line with (thick solid line) and
  without (dashed line) resonant scattering in the
  Perseus cluster. Resonant scattering suppresses the line intensity
  in the core and redistributes line photons to larger radii. Pure
  thermal broadening and a flat radial abundance profile were assumed
  in this calculation. Adapted from \citet{2004MNRAS.347...29C}.
\label{fig:sbprof}
}
\end{figure}

\begin{figure}
\centering
\begin{minipage}{0.49\textwidth}
\includegraphics[trim= 1mm 5cm 20mm 2.1cm,
  width=1\textwidth,clip=t,angle=0.,scale=0.9]{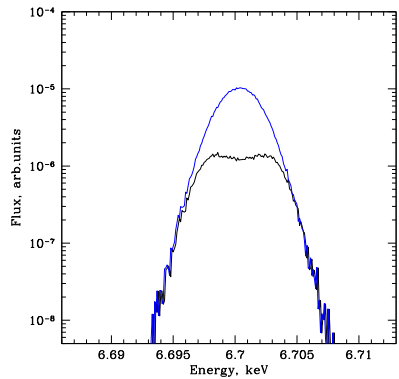}
\end{minipage}
\caption{Simulated spectra of the 6.7 keV line emerging from the core of
  a cluster. Without scattering the line has a Gaussian shape (top blue curve). Due to the large optical depth in the center of the line, the photons are scattered away from the line of sight, suppressing the line emission close to the center. Adapted from \citet{2010SSRv..157..193C}.
\label{fig:line_prof}
}
\end{figure}

The largest optical depth is associated with the densest central part of the cluster. The photons born in this central part are scattered from the line of sight. As a result, the line flux towards the cluster center diminishes,  producing a flattened (see Fig.\ref{fig:sbprof}) surface
brightness distribution \citep{1987SvAL...13....3G}. Due to the conservative nature of resonant
scattering, the photons removed from the line of sight towards the cluster center are
re-distributed to larger projected distances. This effect causes a slight increase of the surface
brightness at larger radii, although this excess emission is more difficult to detect.

Since the radial surface brightness of the line can be extracted from the data with moderate spectral resolution, most of the studies use this approach. In the context of clusters and massive elliptical galaxies the resonant scattering has been most often considered in relation with the ``abundance hole'' or ``abundance anomalies'' in the central regions of clusters \citep[e.g.][]{1998ApJ...499..608M,1999AN....320..283A,1999MNRAS.310..483B,2001A&A...365L.181B,2001ApJ...548..141D,2001ApJ...550L..31M,2001PASJ...53..595E,2002ApJ...579..600X,2004ApJ...600..670G,2006MNRAS.370...63S,2015MNRAS.447..417P}.

Another group of studies was more focused on constraining the turbulent velocities using the line ratios as a proxy to the turbulent broadening of the resonant line \citep[e.g.][]{2004MNRAS.347...29C,2009MNRAS.398...23W,2011AstL...37..141Z,2012A&A...539A..34D,2013MNRAS.433.1172S,2016MNRAS.461.2077P,2017MNRAS.472.1659O}.

Without the resonant scattering the emitted line spectrum and the optical depth of the line are closely related:
\be
F(E)\propto \tau(E)=\tau_{RS} \times e^{-\frac{\left (E-E_{0} \right )^2}{\Delta E_D^2}}.
\ee
Note that this expression ignores Lorentzian wings of the lines and assumes that the line-of-sight gas velocity distribution can be approximated by a Gaussian. Given that the optical depth peaks at the center of the line and goes down quickly in the wings, only the core of the line suffers from the resonant scattering, while the photons in the wings freely escape without any scattering. As a result, the line profile is modified, forming a depression in the center \citep{1987SvAL...13....3G}, as shown in Fig.\ref{fig:line_prof}. For the first time, such modification in a galaxy cluster has been found in the {\it Hitomi} data \citep[][see \S\ref{sec:hitomi} below] {2018PASJ...70...10H}.

\begin{table*}
\scriptsize
\centering
\caption{{\bf Top:} the list of lines in the 5 -- 10 keV energy band in the Perseus cluster with the largest optical depths. {\bf Bottom:} same for the elliptical galaxy NGC~4636 in the 0.5 -- 1 keV band. Only lines with the optical depth $>$ 0.1 are shown. The optical depth is calculated assuming zero Mach number and isotropic turbulence with the Mach number 0.2 (0.25) in Perseus (NGC~4636). The Mach number is defined as $M=V_{1,turb}/c_s$, where $c_s$ is the sound speed.}
\begin{tabular}{@{}l@{}lccccc}
\hline
{\bf Perseus Cluster} & & \\
\hline
Ion & $E$, keV & Transition & $f$ &$\tau_{\rm M=0}$ & $\tau_{\rm M=0.2}$\\
\hline
Fe XXIV  & 6.66188 & 1s$^2$2s($^2$S$_{1/2}$) - 1s2s2p($^2$P$_{3/2}$) & 0.4892 & 0.216  & 0.080\\
Fe XXV   & 6.66755 &  1s$^2$($^1$S$_0$) - 1s2p($^3$P$_1$) & 0.0579 & 0.170 & 0.063\\
Fe XXV   & 6.70041 &  1s$^2$($^1$S$_0$) - 1s2p($^1$P$_1$) & 0.7192& 2.100 & 0.780 \\
Fe XXVI  & 6.97307 & 1s($^2$S$_{1/2}$) - 2p($^2$P$_{3/2}$) & 0.2731& 0.147 & 0.054\\
Fe XXV   & 7.88153 & 1s$^2$($^1$S$_0$) - 1s3p($^1$P$_1$) & 0.1369& 0.340 &  0.126\\
Fe XXV & 8.29548 & 1s$^2$($^1$S$_0$) - 1s4p($^1$P$_1$)& 0.0478 & 0.113 & 0.042 \\
\hline
\hline
{\bf NGC 4636} & & \\
\hline
Ion & $E$, keV & Transition & $f$ & $\tau_{\rm M=0}$ & $\tau_{\rm M=0.25}$\\
\hline
O VIII &   0.65349 & 1s($^2$S$_{1/2}$) - 2p($^2$P$_{1/2}$)   & 0.1385 & 0.409  & 0.210\\
O VIII   & 0.65368 & 1s($^2$S$_{1/2}$) - 2p($^2$P$_{3/2}$) & 0.2771 & 0.819 &  0.420\\
Ne IX & 0.92200 & 1s$^2$($^1$S$_0$) - 1s2p($^1$P$_1$) & 0.7417 & 0.348 & 0.163\\
Fe XVII &  0.72714 & 2s$^2$2p$^6$ ($^1$S$_0$) - 	2s$^2$2p$^5$3s(3/2,1/2)$_1$	& 0.1257 & 0.340 & 0.104 \\
Fe XVII  & 0.81243 & 2s$^2$2p$^6$($^1$S$_0$) - 2s$^2$2p$^5$3d($^3$D$_1$) & 0.6391 & 1.545 &  0.471\\
Fe XVII  & 0.82579 & 2s$^2$2p$^6$($^1$S$_0$) - 2s$^2$2p$^5$3d($^1$P$_1$)& 2.4938 & 5.933 & 1.808\\
Fe XVII  & 0.89681 & 2s$^2$2p$^6$($^1$S$_0$) - 2s2p$^6$3p($^1$P$_1$) & 0.2869 & 0.628 & 0.192 \\
Fe XVIII & 0.85306 & 2s$^2$2p$^5$($^2$P$_{3/2}$) - 2s$^2$2p$^4$3d($^2$F$_{5/2}$)  & 0.2045 & 0.417 & 0.127\\
Fe XVIII & 0.86262 & 2s$^2$2p$^5$($^2$P$_{3/2}$) - 2s$^2$2p$^4$3d($^2$D$_{5/2}$) & 0.3083 & 0.622 & 0.190\\
Fe XVIII & 0.86970 & 2s$^2$2p$^5$($^2$P$_{3/2}$) - 2s$^2$2p$^4$3d($^3$S$_{1/2}$)	& 0.2276 & 0.456 & 0.139\\
Fe XVIII & 0.87264 & 2s$^2$2p$^5$($^2$P$_{3/2}$) - 2s$^2$2p$^4$3d($^2$P$_{3/2}$) & 0.6116 & 1.219 & 0.372 \\
Fe XVIII & 0.87264 & 2s$^2$2p$^5$($^2$P$_{3/2}$) - 2s$^2$2p$^4$3d($^2$D$_{5/2}$)& 0.9295 & 1.854 & 0.565\\
Fe XIX  &  0.91718 & 2s$^2$2p$^4$($^3$P$_2$) - 2s$^2$2p$^3$3d(5/2,5/2)$_3$ & 0.7438 & 0.854 & 0.260 \\
Fe XIX  &  0.91861 & 2s$^2$2p$^4$($^3$P$_2$) - 2s$^2$2p$^3$3d(5/2,3/2)$_3$ & 0.3781 & 0.434 & 0.132\\
Ni XIX   &  0.99706 & 2s$^2$2p$^6$ ($^1$S$_0$) - 2s$^2$2p$^5$3d(1/2,3/2)$_1$ & 2.5214 & 0.324 & 0.097\\
\hline
\label{tab:lines}
\end{tabular}
\end{table*}

\subsection{Modeling resonant scattering}

Spectral and spatial distortions due to the resonant scattering have been estimated analytically in a small-$\tau$ limit for isothermal gas \citep{1987SvAL...13....3G,2002MNRAS.333..191S}. For real objects, it is necessary to perform radiative transfer simulations, which take into account gas temperature gradient and multiple scatterings of photons. The typical optical depth in individual lines is $\lesssim$ few in galaxy clusters, groups and elliptical galaxies (see Table \ref{tab:lines}). Therefore, radiative transfer effects can be efficiently modeled following a Monte Carlo approach. In simulations, the act of scattering can be modeled in full details. First, a photon with the energy $E$ is absorbed by an ion, moving with such velocity that in the frame of the ion, the photon energy is equal to the energy $E_0$ of the line, i.e.,
\be
E=E_0\displaystyle\left[1+\frac{({\bf V_{\rm ion}m})}{c}\right],
\ee
where ${\bf V}_{\rm ion}$ and $\bf m$  are the ion velocity and  the direction of the photon propagation in the cluster frame, respectively\footnote{here we assume that ${V}_{\rm ion}\ll c$}. Given that the above expression depends only on the scalar product $({\bf V}_{\rm ion}{\bf m})$, there are no constraints on the two components of $\bf V_{\bf ion}$ that are perpendicular to $\bf m$. These two components are drawn from a Gaussian distribution with the width set by the Doppler line width. The photon is scattered in the frame moving with ion (according to the relevant isotropic and/or dipole phase function) and transformed back to the cluster frame. After each scattering the weight, initially assigned as unity to each photon, is reduced by a factor of $(1-e^{-\tau})$, where $\tau$ is the optical depth in the direction of photon propagation. The scattering process repeats until the weight drops below the set minimal value.

Two independent approaches of simulating resonant scattering have been implemented: using a Geant4\footnote{http://geant4.cern.ch} toolkit for the simulations of the passage of particles through matter \cite[see e.g.][]{2018PASJ...70...10H} and Monte Carlo simulations tailored for X-ray photons propagating through gas-rich atmospheres \citep[e.g.][and references therein]{2002MNRAS.333..191S, 2018PASJ...70...10H}. The latter simulations considered, initially, only individual, optically-thick lines, assumed a spherically-symmetric geometry and an isotropic turbulence \citep{2002MNRAS.333..191S,2004MNRAS.347...29C}. These simulations have been later extended to study the resonant scattering in galaxy clusters taken from cosmological simulations \citep{2010MNRAS.403..129Z}, in anisotropic velocity field \citep{2011AstL...37..141Z} and in a broad range of energies, producing spectral models suitable for direct comparison with observed spectra \citep{2013MNRAS.435.3111Z}.

\begin{figure}
\centering
\includegraphics[trim= 10 0 0 0, width=0.48\textwidth]{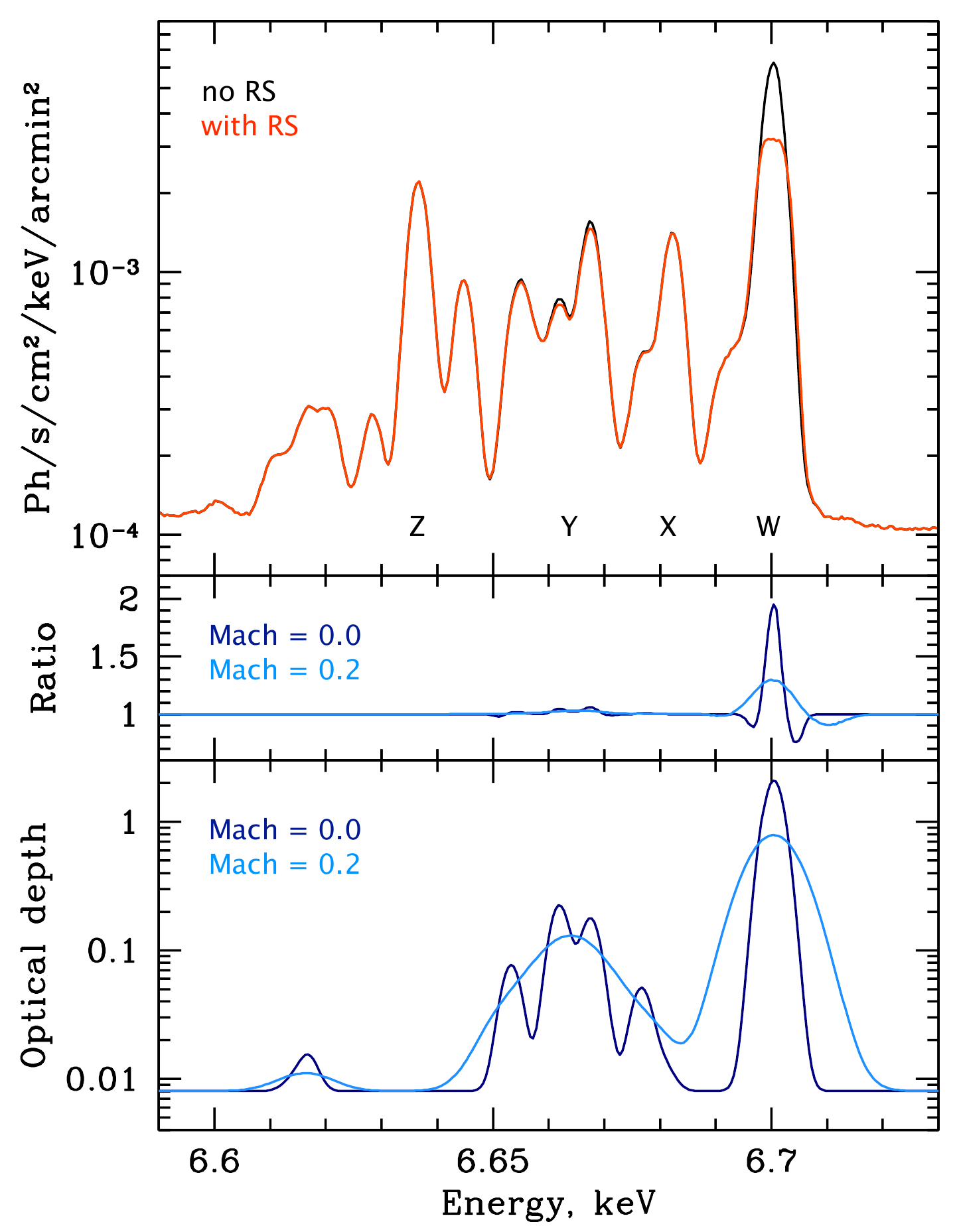}
\caption{Resonant scattering in the 0.5' -- 2' annulus in the Perseus cluster. Only the spectrum of He-like Fe triplet is shown. {\bf Top:} spectrum with (red) and without (black) resonant scattering. {\bf Middle:} ratio of the two models when gas is static (M=0.0, navyblue) and turbulence is present in the gas with a Mach number 0.2 (dodgerblue). {\bf Bottom:} optical depth as a function of photon energy for the two cases of Mach numbers. Adapted from \citet{2013MNRAS.435.3111Z}, updated for the latest AtomDB model (v3.0.9).
\label{fig:perseus_rs}
}
\end{figure}

Taking line emissivities and continuum emission from the latest AtomDB (v3.0.9) model of an optically thin plasma, we updated the results of the resonant scattering simulations in the core of the Perseus cluster \citep[for earlier results see][]{2013MNRAS.435.3111Z}. Fig. \ref{fig:perseus_rs} shows projected spectra with and without resonant scattering in the 0.5' -- 2' annulus, the ratio of these spectra and the optical depth as a function of photon energy. The spectra are shown for the He-like Fe line complex. Middle and bottom panels show the results when gas is static (M=0, navyblue curves) and when turbulence is present in the gas with the Mach number of one-component velocity $M=V_{1,turb}/c_s = 0.2$ (dodgerblue). Several lines in the He-like Fe triplet are affected by scattering. The resonant Fe line (W) at 6.7 keV has the largest optical depth $\tau\sim2$ (M=0) and $\sim 0.8$ (M=0.2). As the result of scattering, the flux in this line is suppressed by a factor of 2 (1.3) when $M=0$ ($M=0.2$) and the wings of the line become stronger (little troughs in the middle panel in Fig. \ref{fig:perseus_rs}). Modeling such distorted line with a Gaussian will effectively give a larger line width compared to the other lines of the same ion. Intercombination lines (Y) in the Fe triplet are slightly affected by the resonant scattering as well. However, line distortions disappear even for a relatively low turbulence.

\begin{figure*}
\centering
\includegraphics[trim= 0 0 0 0, width=0.98\textwidth]{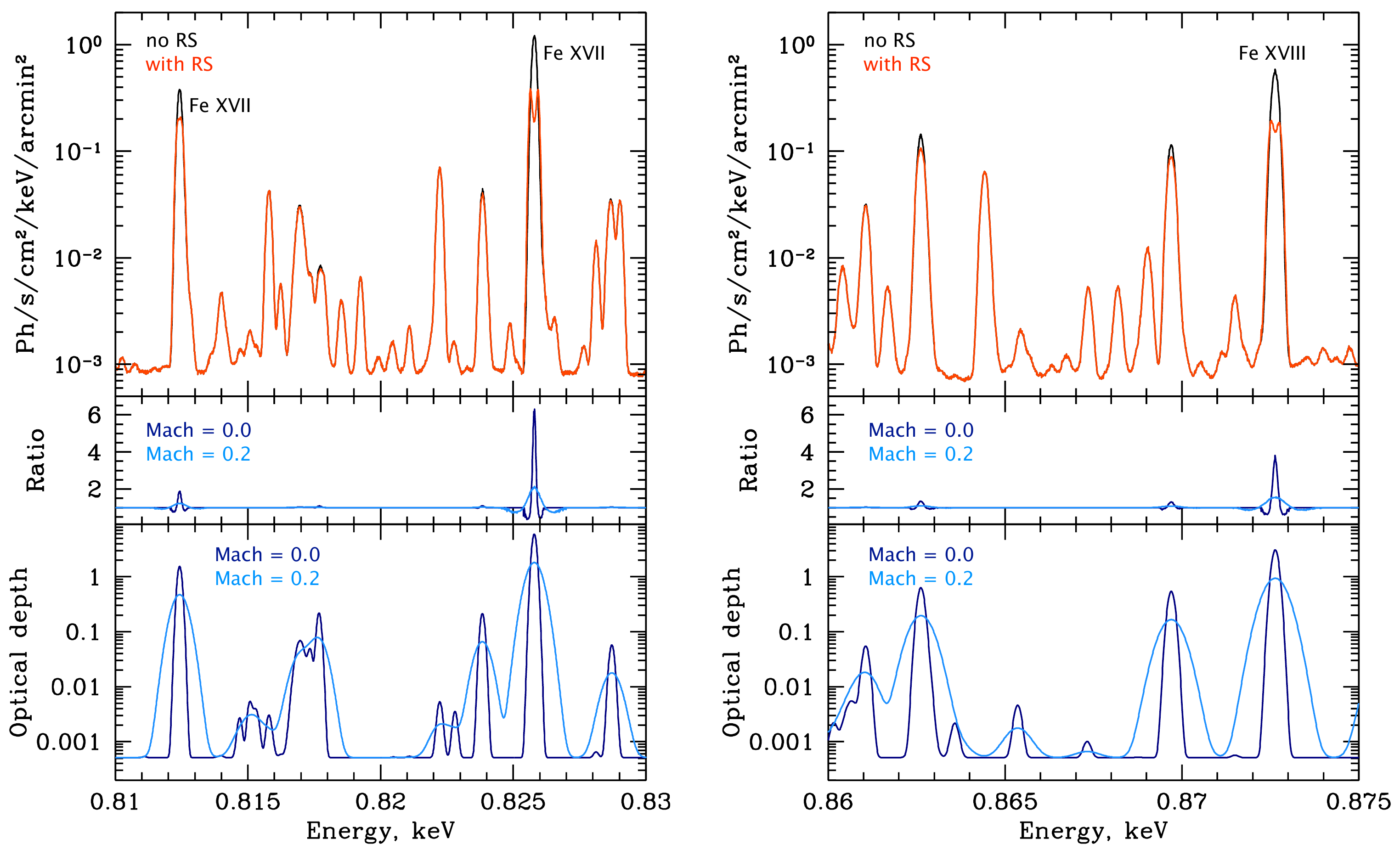}
\caption{Resonant scattering in the inner kpc region in NGC~4636. Only the portions of spectra containing lines with the largest optical depth are shown. Color coding is the same as in Fig. \ref{fig:perseus_rs}. 
\label{fig:ngc4636_rs}
}
\end{figure*}

Fig. \ref{fig:ngc4636_rs} shows the resonant scattering effect in a massive elliptical galaxy NGC~4636. The gas in NGC~4636 is significantly cooler than in Perseus, the average temperature is $<$ 1 keV. For such temperatures, the strongest emission lines are the lines of the Ne-like Fe at 15\AA\, and 17\AA. The optical depth of the strongest resonant line at 0.826 keV is $\sim$ 6 (Table \ref{tab:lines}), almost three times larger than the optical depth in the W line in Perseus. Resonant scattering suppresses the flux in this line by a factor of more than 6 and 2 for the Mach numbers 0 and 0.25, respectively (Fig. \ref{fig:ngc4636_rs}). Several other lines strongly affected by the scattering, including the Fe XVIII line at 0.873 keV with the flux suppression by a factor of $\sim$ 4 when $M=0$ and the Fe XVII line at 0.812 keV with the suppression by a factor of $\sim$ 2 (Fig. \ref{fig:ngc4636_rs}).

\subsection{Observing resonant scattering}
\label{sec:hitomi}
Spectral distortions due to the resonant scattering effect have been searched in many objects. Using high-resolution RGS/XMM-Newton spectra, a clear evidence of the effect in the Fe~XVII line at 15.01\AA\, has been found in the bright and compact cores of several elliptical galaxies, including NGC~4636, NGC~1404, NGC~5813 and NGC~4472 \citep{2002ApJ...579..600X,2003ASPC..301...23K,2009MNRAS.398...23W,2012A&A...539A..34D}. \citet{2016A&A...592A.145A} studied spatial variations of the Fe~XVII optically thin to thick line ratio in NGC~4636, suggesting that the resonant scattering effect is smaller in the southern and stronger in the northern regions of the galaxy. Recently, \citet{2017MNRAS.472.1659O} performed a detailed analysis of the resonant scattering detected with the RGS observations in a sample of 22 nearby, giant elliptical galaxies and, combining the analysis with the Doppler broadening technique, constrained the turbulent velocity in 13 objects. There are observational indication of the resonant scattering in a cooler gas component with the temperature $\sim 2\cdot 10^6$ K, which is traced by the OVII emission lines. In several elliptical galaxies, the observed OVII resonant-to-forbidden line ratio is decreased, which could be due to the resonant scattering or charge exchange \citep{2016MNRAS.461.2077P}. The same authors found that OVII is mainly detected in galaxies, where the Fe~XVII resonant line is suppressed by the resonant scattering.

Despite many attempts to detect the resonant scattering effect in the Perseus cluster \citep[e.g.][]{1998ApJ...499..608M,2001PASJ...53..595E,2004MNRAS.347...29C} and other clusters \citep{1999ApJ...519L.119K,2000AdSpR..25..603A,2001ApJ...550L..31M,2002A&A...391..903S,2006MNRAS.370...63S}, the conclusions remained controversial. The energy resolution of the CCD-type spectrometers is not sufficient to distinguish between spectral distortions due to the resonant scattering, variations of gas temperature and metallicity and gas motions \citep{2013MNRAS.435.3111Z}. Also, flux suppression due to the resonant scattering in the resonant Fe XXV line is often weaker than in the resonant Fe XVII line in elliptical galaxies. Recent RGS observations of the Perseus cluster show the Fe XVII forbidden-to-resonant line ratio is significantly higher than predicted from optically thin plasma, suggesting that the resonant scattering might be important in the cluster core \citep{2016MNRAS.461.2077P}.

\begin{figure*}
\centering
\includegraphics[trim= 0 0 0 0, width=0.98\textwidth]{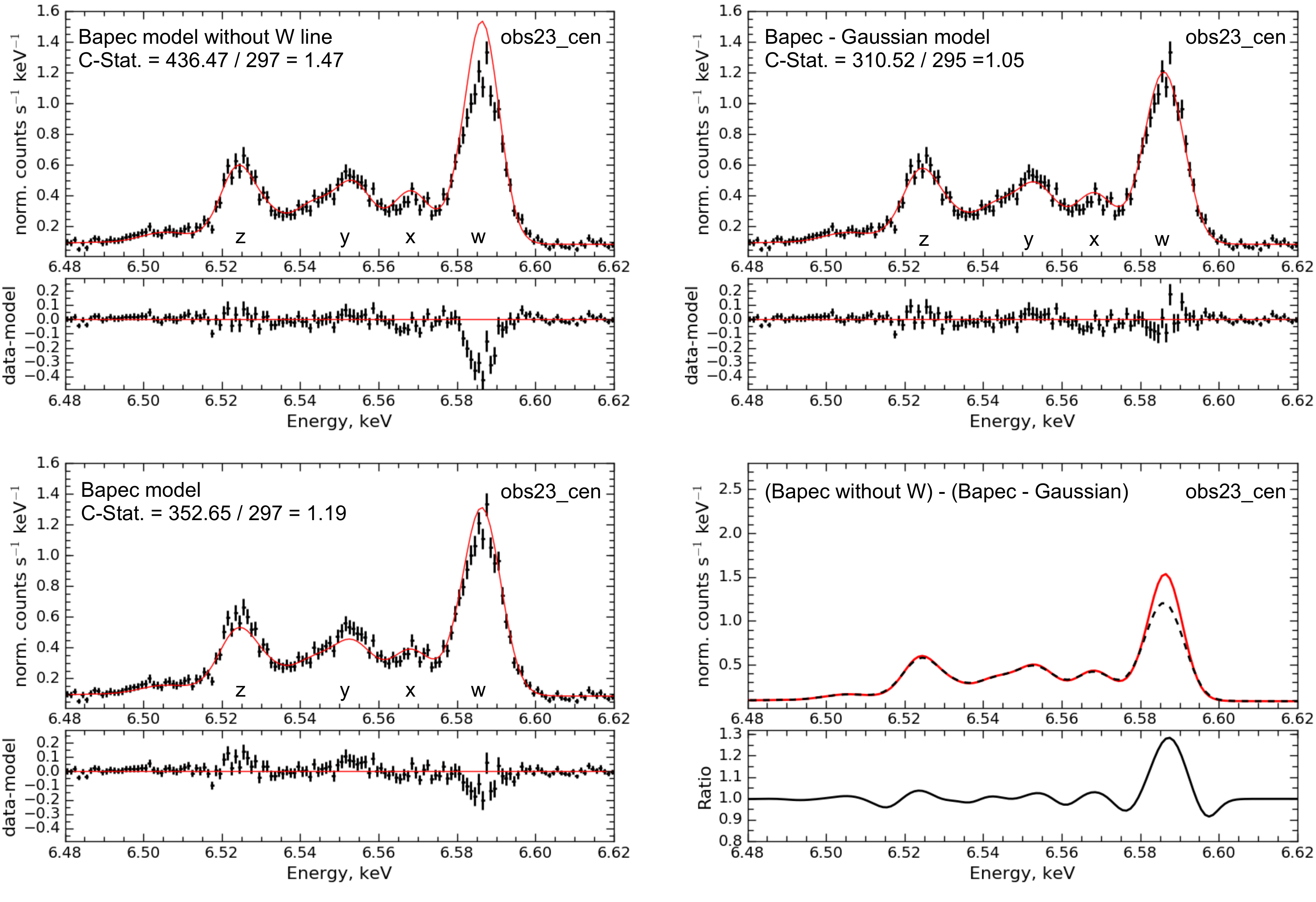}
\caption{Observed flux suppression in the strongest line of He-like Fe XXV (w) in the Perseus Cluster observed in inner $\sim$ 20 kpc region. Black points show the {\it Hitomi} data; red lines in the corresponding panels show the best-fitting models. {\bf Top left:} the spectrum is fitted with the bapec model from AtomDB, excluding the w line from the data; {\bf top right:} the same spectrum is fitted with the bapec model and a Gaussian component centered at the energy of the w line, the normalization of the Gaussian model is allowed to be negative; {\bf bottom left:} the same spectrum fitted with the bapec model. The comparison of the models from the top left (solid) and right (dashed) panels is shown in the bottom right panel. Adapted from \citet{2018PASJ...70...10H}
\label{fig:hitomi}
}
\end{figure*}

The resonant scattering controversy in the Perseus cores has been recently resolved with the {\it Hitomi} observations. Due to the superb energy resolution of 5 eV at 6 keV of the soft X-ray Spectrometer on-board {\it Hitomi}, the individual spectral lines have been resolved \citep{2018PASJ...70...12H} and the suppression of the resonant Fe XXV line has been measured robustly \citep{2018PASJ...70...10H}. Fig. \ref{fig:hitomi} shows the He-like Fe triplet observed in the Perseus inner $\sim$ 20 kpc region and modeled with an optically thin plasma model from AtomDB, accounting for the flux suppression in the resonant W line. The W line flux is suppressed by a factor of $\sim$ 1.3 in the innermost $\sim$ 40 kpc and by a factor of $\sim$ 1.15 in the region $\sim$ 30 -- 100 kpc away from the cluster center. Both, the amplitude of the flux suppression and the variations of the suppression with the projected distance from the cluster center, are in excellent agreement with the theoretical predictions of the resonant scattering. Also, an additional broadening \footnote{when the line is approximated with a Gaussian} of the resonant line compared to other lines from the same ion has been observed, supporting a non-Gaussian shape of the W line.

\subsection{Uncertainties in the atomic physics}

The resonant scattering is measured by comparing the observed line ratios with the models,
which introduces systematic uncertainties from the underlying atomic physics. The uncertainties mainly 
originate in two parameters: the oscillator strengths of the resonant lines, and the
model intensity ratio of the resonant-to-forbidden lines.

A systematic estimate of the atomic uncertainties was given in \citet{2018PASJ...70...10H},
for the study of the Perseus spectrum ($k$T $\approx$ 4~keV) taken with the {\it Hitomi} satellite.
The oscillator strength of the Fe XXV He$\alpha$ resonant transition was found to be well determined,
introducing a systematic error of $<5$\%. The resonant-to-forbidden line ratio has a more
complex nature. The uncertainty is contributed by three components, the errors in the line formation kinetics, 
errors associated with the unresolved satellite lines, and errors from the possible 
charge exchange recombination. The first component can be assessed with the 
ground laboratory measurement at the electron beam ion trap (EBIT) device. By comparing
the current plasma code (AtomDB/APEC and SPEX) with the EBIT line ratio for a 4~keV Maxwellian plasma reported in \citet{2012CaJPh..90..351G}, the error associated with the line formation was found to
be $\sim 6$\%. The second component was estimated by comparing the unresolved satellite lines
obtained with the AtomDB/APEC, SPEX, and the ground experiments. The total error on the Fe XXV
resonant-to-forbidden line ratio is about 4\% for a 4~keV plasma. As for the third component,
the recombination from charge exchange between Fe XXVI and neutral atom can suppress the 
resonant-to-forbidden line ratio by $\sim 6$\% in the core of the Perseus cluster. A detailed
review on the charge exchange process will be given in \S\ref{sec:4}.

Combining the errors associated with the oscillator strength and resonant-to-forbidden line ratio,
the total systematic uncertainty for the resonant scattering in the Perseus cluster is about 10\%.
For the cooler systems, such as elliptical galaxies, the atomic errors can be larger than those for the hot clusters. As shown in \citet{2017MNRAS.472.1659O}, the error from the line formation kinetics
on the Fe XVII resonant-to-forbidden line ratio is $\sim 15$\%. Future theoretical calculations
and targeted experiments providing better agreement on the atomic properties are needed to match 
the precision of the existing and future high spectral resolution X-ray missions.

\section{Non-thermal effects}
\label{sec:2}

\begin{figure}[!htbp]
\centering
\resizebox{0.9\hsize}{!}{\includegraphics[angle=-90]{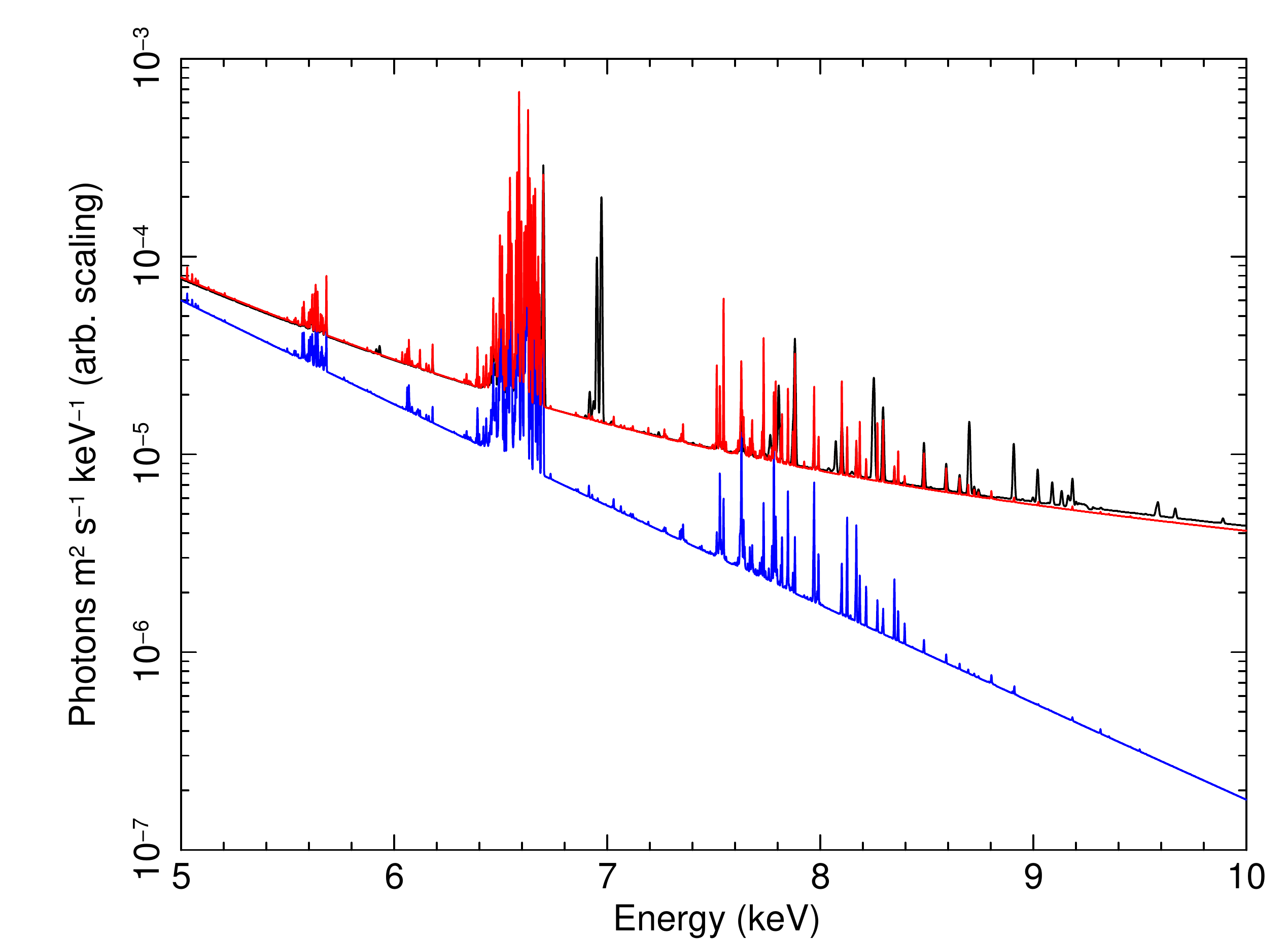}}
\caption{Blue curve: collisional ionisation equilibrium (CIE) $kT=1$~keV spectrum; Black curve: sum of a $kT=1$~keV CIE spectrum and a $kT=10$~keV CIE spectrum, with the hotter component having 1\% of the emission measure of the colder component; Red curve: spectrum of a plasma with $kT=1$~keV thermal electron distribution plus a 1\% contribution from a $kT=10$~keV electron distribution.
}
\label{fig:nonth}
\end{figure}

Non-thermal electrons can be produced in clusters of galaxies in various ways. 
Near the interface of cold and hot gas, thermal electrons from the hot gas may penetrate into the cold gas and constitute there a population of super-thermal electrons. 
This process only works when the magnetic field geometry allows these electrons to pass. 
Acceleration of electrons is perhaps the most attractive mechanism to produce non-thermal electrons. 
This can happen near magnetic neutral points during magnetic reconnection by the $v\times B$ electric fields, or likely more frequently as a product of shocks.

Observing these non-thermal electrons offers direct insight into the physical processes that create them and hence in the physics of the cluster gas.
One may distinguish three ways in which non-thermal electrons affect the X-ray spectrum of hot gas \citep{2009A&A...503..373K}: 1) the production of a non-thermal bremsstrahlung tail to the continuum spectrum, 2) a shift in the ionisation balance towards higher ionisation, and 3) modified line intensities due to resonant processes.

Non-thermal bremsstrahlung tails can be detected by searching for excess continuum emission at high energies. We refer to \citet{2008SSRv..134...71R} for a review on that process. The effects on the ionisation balance are in general not very large, but should be taken into account, unless the non-thermal electrons are short-lived, in which case they might affect the instantaneous spectrum, but the plasma may not have had enough time to adjust. The SPEX package \citep{1996uxsa.conf..411K} allows to treat both cases.

Non-thermal electrons also affect line ratios. This happens for all resonant processes, like dielectronic recombination or collisional excitation resonances. Normal excitation processes, like collisional excitation by the hot plasma, have different cross sections as a function of energy, but are possible for all electrons in the plasma. Hence they are usually dominated by the bulk of the (thermal) electrons. Resonances, on the other hand, are only possible for a specific electron energy $E_r$. Resonant contributions to the line intensities are thus determined by the relative contribution of the electron distribution at $E_r$ versus the bulk of the thermal electrons. For each transition, this ratio is different and thus dielectronic satellite lines offer an excellent opportunity to demonstrate the presence and quantify the amount of non-thermal electrons \citep{1979MNRAS.189..319G}.

Technically, the presence of non-thermal electrons in the spectrum is best accounted for by writing the electron distribution as the sum of Maxwellians. This is because the bulk of all atomic processes are characterised by rates that are averaged over a thermal electron distribution. The SPEX package \citep{1996uxsa.conf..411K} has implemented the treatment of non-thermal electrons in this way. Recently, \citet{2015ApJ...809..178H} have provided a convenient parameterisation of the so-called kappa-distribution in terms of a sum of Maxwellians.

Care should be taken when interpreting spectra with non-thermal electrons. Fig.~\ref{fig:nonth} illustrates this. The red curve is a spectrum of a 1~keV plasma with an additional non-thermal electron contribution of 1\% with a 10 times higher energy. Compared with a simple 1~keV thermal spectrum (blue curve) the most striking feature is the non-thermal bremsstrahlung tail most visible at the higher energies. In order to demonstrate that this is the spectrum of a plasma with non-thermal electrons, sensitive lines should be used but in addition it should be shown that the spectrum is not simply the sum (or superposition) of two separate plasmas with different temperatures (1 and 10 keV) and emission measures (the hotter component 1\% of the colder component); this is the black curve.

It is seen immediately that the main difference is the absence of the Fe~XXVI Lyman$\alpha$ lines near 7~keV in the red spectrum, as well as some other lines from the hot plasma. In the black, two-temperature spectrum with Lyman$\alpha$ line is solely produced by the hot gas. In the red, non-thermal spectrum there are simply no Fe~XXVI ions so the Lyman$\alpha$ cannot be produced at all.

The most sensitive lines for a 1~keV plasma with a non-thermal tail are the the Fe~XXVI Lyman$\alpha$ lines and several of the stronger Fe XVII lines in the 0.7--0.8~keV range. These lines are weaker than for a comparable two-temperature plasma. On the other hand, the Fe XXIV line at 6.662~keV is the best line that becomes stronger.

For a 2~keV plasma, the Fe~XXVI Lyman$\alpha$ lines are the best diagnostic with weakening due to the non-thermal electrons, and the higher Rydberg lines of Fe~XXV with strengthening of the intensity. 

For a 4~keV plasma, however, the Fe XXVI Lyman$\alpha$ lines become stronger when non-thermal electrons are present, compared with a two-temperature plasma. This is mainly due to the fact that the excitation rate for this line is low for $\sim$40~keV electrons.

%$\rightarrow$ Kaastra, Gu

%2.1 electron-impact excitation (maybe a list of non-thermal-sensitive lines) \\
%2.2 other possible non-thermal effects 

\section{Non-equilibrium effects}
\label{sec:3}

%$\rightarrow$ Yamaguchi
%P1: physics basics of non-equilibrium ionization and spectral features. 
%P2: The physical link with cluster accretion shocks and merger shocks.
%References: Fujita et al. 2008, Akahori \& Yoshikawa 2010, Wong, Sarazin, and Ji 2011, 
%and Inoue et al. 2016
%P3: perspectives for future instruments.

\bigskip

Because of an extremely low density ($\lesssim 10^{-3}$ cm$^{-3}$), 
the typical mean free path of particles in the ICM is long. In such condition, 
thermal equilibration between electrons and ions is achieved slowly with 
the timescale given by 
\begin{equation}
 t_{\rm ei} \sim 7 \times 10^7 
 \left( \frac{n_{\rm e}}{10^{-3}\,{\rm cm}^{-3}} \right)^{-1} 
 \left( \frac{T_{\rm e}}{5 \times 10^7\,{\rm K}} \right)^{3/2}
\end{equation}
(Spitzer 1962). 
Although this timescale is shorter than the typical cluster age,  
thermal non-equilibrium could exist in the shock front of 
merging clusters or cluster outskirt. 
In addition, the so-called non-equilibrium ionization (NEI) effect, 
where the ionization degree of ions departs from that in an 
equilibrium plasma, is also expected in such regions 
\citep{2010PASJ...62..335A, 2011ApJ...727..126W}.

There has been some observational tests for NEI conditions in the ICM. 
\citet{2008PASJ...60.1133F} analyzed Suzaku data of the Ophiuchus cluster and 
found that the observed line ratio of Fe Ly$\alpha$ to He$\alpha$ emission  
is consistent with that predicted by the electron temperature determined 
from the bremsstrahlung continuum. Similarly, no positive evidence for 
NEI plasma was found in several merging clusters 
(e.g., Abell 2146: \citealt{2012MNRAS.423..236R}, CIZA\,J2242.8+5301). 
Recently, Suzaku observations of Abell 754 has revealed a signature 
of the NEI \citep{2016PASJ...68S..23I}. They found that spectra from the 
highest-temperature region are well characterized by an NEI plasma 
 with $n_e t \sim 4 \times 10^{11}$\,cm$^{-3}$\,s. 
From this result, they estimated the elapsed time from the shock 
heating to be 0.36--76\,Myr, consistent with the traveling time 
of a shock to pass through that region.

More recently, high-resolution {\it Hitomi}/SXS spectra of the Perseus cluster 
provided the first opportunity to measure the ion temperature in the ICM 
using the width of well-resolved emission lines \citep{2018PASJ...70...11H}. 
The line-of-sight velocity dispersion due to an isotropic thermal motion of 
ions is given by $\sigma_{\rm th} = \sqrt{kT_{\rm ion}/m_{\rm ion}}$, 
which depends on the mass of nuclei. On the other hand, the kinetic 
broadening due to the turbulence motion $\sigma_{\rm v}$ is common among 
the elements. Therefore, we can separate the contributions of $\sigma_{\rm th}$ 
and $\sigma_{\rm v}$ by measuring the widths of different elements.   
\citet{2018PASJ...70...11H} used emission lines of Si, S, Ca and Fe, 
and revealed that the ion temperature is consistent with the electron temperature. 
Future observations with better sensitivity to lighter elements (e.g., O, Ne) will 
allow us to investigate the non-equilibrium effects (both thermal and ionization) 
in a number of clusters.

\section{Charge exchange}
\label{sec:4}

%$\rightarrow$ Gu

\begin{figure*}[!htbp]
\centering
\resizebox{0.9\hsize}{!}{\includegraphics[angle=0]{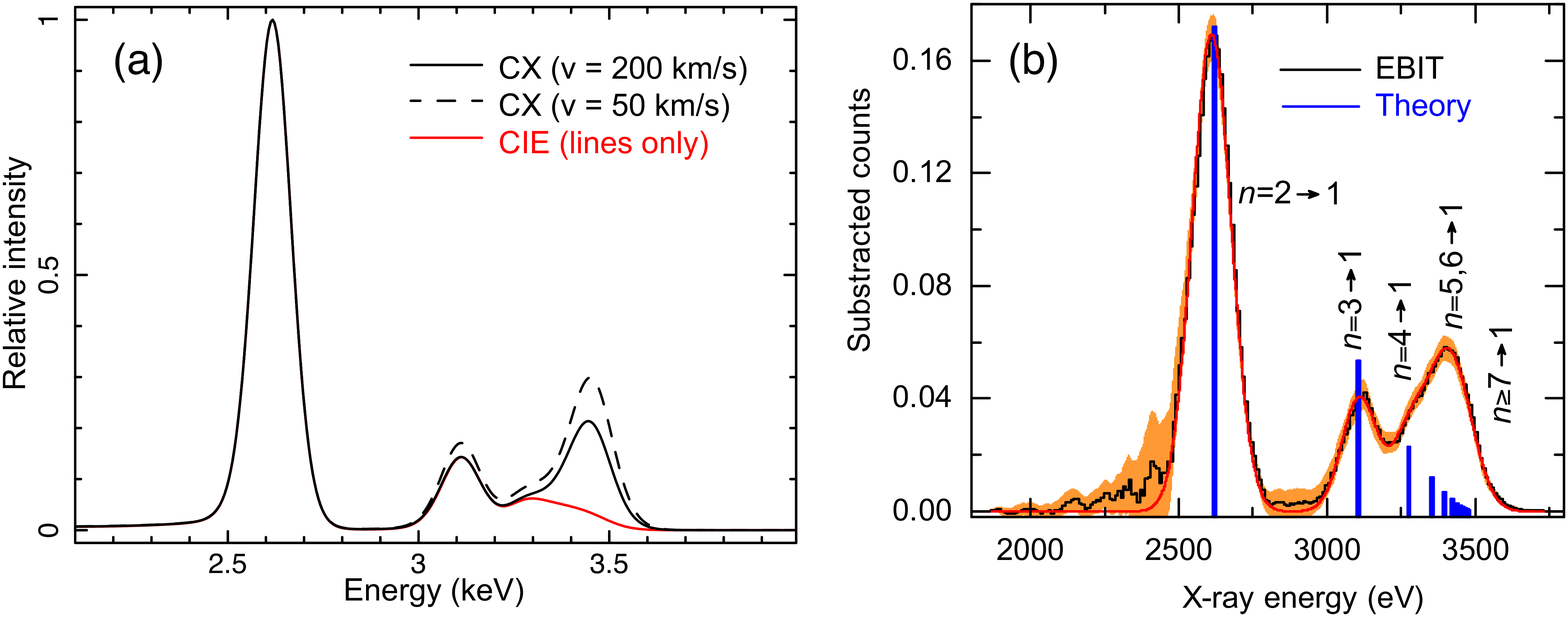}}
\caption{(a) Model spectrum for charge exchange of bare sulfur with atomic hydrogen, based on
the plasma model of \citet{2016A&A...588A..52G}, compared
with the thermal line emission of hydrogen-like sulfur. (b) Charge exchange emission of bare sulfur
with CS$_{2}$ measured in the electron beam ion trap experiment \citep{2016ApJ...833...52S}. Theoretical predictions of the high-Rydberg-state energies are shown by blue vertical lines.
\vspace{0.5cm}
}
\label{fig:cx}
\end{figure*}

Charge exchange (CX) occurs when neutral and ionized particles mix with each other. One or some electrons can be stripped
from the cold atom and bind instead to the hot ion, for the latter often has a deeper ionization potential. The cross
section of the process is usually at least two orders of magnitude larger than the free-electron-impact cross section
at the keV energies, making it very effective for ionizing neutral atoms, and for reducing the ionization degree
of the hot plasma. In the astrophysical conditions where CX is allowed, it could easily be the dominating radiation
process.

The captured electron relaxes to the ground state via line emission. Atomic theory suggests that the electron is likely to 
drop onto a high Rydberg state with large principle quantum number $n$, creating signature line radiation 
which appears at a shorter wavelength than those from collisional excitation, but at a longer wavelength 
than the radiation recombination edge. It also gives a high forbidden-to-resonance ratio for the He$\rm \alpha$ triplet. 
Recently, accurate and large-scale theoretical modeling of CX became available \citep{2001ASPC..247....3S}, giving rise to new plasma models in both the AtomDB and SPEX codes \citep{2012AN....333..301S, 2016A&A...588A..52G}. Benchmarking these models to the laboratory experiments achieves reasonable agreement 
\citep{2016ApJ...833...52S, 2018ApJ...852....7C}.

Astrophysical observation of the CX phenomena at the X-ray wavelength is traced back to 1996, when
comets were discovered as a new class of X-ray sources \citep{1996Sci...274..205L}. It was soon found that the 
CX between solar wind and comet atmosphere produces the observed soft X-ray \citep{1997GeoRL..24..105C}. The presence 
of solar wind CX from the planet atmosphere \citep{2006A&A...451..709D, 2007A&A...463..761B} and 
from the heliosphere \citep{2004ApJ...610.1182S} has also been observationally established. For the extrasolar sources,
theoretical studies predicted CX at the cold cloud surface in the hot interstellar medium \citep{2009SSRv..143..427L} and 
intracluster medium \citep{2004A&A...422..391L, 2015A&A...584L..11G}. Real observations reveals both successes and 
challenges. There are a growing number of reports of CX-like He$\rm \alpha$ triplet and/or high 1$s$-$np$ fluxes seen
in the X-ray spectra of supernova remnants (e.g., Cygnus loop, \citealt{2011ApJ...730...24K, 2014ApJ...787L..31C, 2015MNRAS.449.1340R}), star-bursting galaxies (e.g., M~82, \citealt{2011MNRAS.415L..64L, 2011PASJ...63S.913K, 2014ApJ...794...61Z, 2016MNRAS.458.3554C}), elliptical galaxies \citep{2016MNRAS.461.2077P}, and clusters of galaxies \citep{2011MNRAS.417..172F, 2015MNRAS.453.2480W, 2018A&A...611A..26G}. However, 
the observed line emission is either too dim \citep{2017ApJ...837L..15A}, or is also explained by other competing processes 
\citep{2016A&A...594A..78G}.

An intriguing application of CX in clusters of galaxies is to explain the possible line feature at the X-ray energy of 
3.5 keV. In 2014, two groups claimed to have detected an unknown line feature at around 3.5 keV in the stacked CCD spectra of
a large cluster sample \citep{2014ApJ...789...13B, 2014PhRvL.113y1301B} . The line significance is on the margin, even though the authors managed to enhance the contrast by 
removing the continuum and known atomic lines from the spectra. The reports attract enormous attention, giving rise to a tide
of the publications attempting to explain the possible nature of the unidentified line. The main theoretical focus
is towards the possibility that the line is created by the decay of a hypothetical dark matter particle candidate.
Recently, \citet{2015A&A...584L..11G} pointed out an alternative interpretation: the observed line arises from CX between the
fully striped sulfur ion and atomic hydrogen, populating a high Rydberg state of the product H-like sulfur ion (Fig.~\ref{fig:cx}). The resulting
transitions cannot be found in the present atomic database for thermal plasmas. The two ingredients, the neutral and fully ionized particles, are found to co-exist in galaxy clusters \citep{2001AJ....122.2281C}. The theoretical calculation of \citet{2015A&A...584L..11G} has been verified with an electron beam ion trap experiment \citep{2016ApJ...833...52S}.

\citet{2014ApJ...789...13B} and others reported an anomalously high 3.5 keV line flux in the Perseus cluster compared to
the other clusters. The line was expected to be detected with the {\it Hitomi} SXS, which is the only instrument capable
of fully resolving this feature. By analyzing the {\it Hitomi} data of the Perseus center, \citet{2017ApJ...837L..15A} 
reported the non-detection of the line at the expected flux, inconsistent with the CCD results at $>99$\% confidence 
level. Instead, the {\it Hitomi} spectrum did reveal a hint of the charge exchange feature at 3.44 keV (S XVI) and  
8.63 keV (Fe XXV), although the significances are still low (1.6$\sigma$ for sulfur and 2.4$\sigma$ for iron) with the current data \citep{2018PASJ...70...12H}. As noted in \citet{2017ApJ...837L..15A}, the central energy of the potential S XVI
CX line is about 2.6$\sigma$ away from the best-fit energy of the unknown 3.5~keV line for the {\it XMM-Newton} MOS detection
\citep{2014ApJ...789...13B}. Therefore, it is unclear whether the 3.5~keV line can be fully explained by the sulfur charge exchange,
or it still requires other ingredients.

Besides the sulfur and iron, by examining the stacked {\it XMM-Newton} RGS data for a sample of 21 cool-core galaxy 
clusters, \citet{2018A&A...611A..26G} reported the signature of O VIII charge exchange at about 0.84~keV with a 
significance of 2.8$\sigma$. This result might affect the ICM abundance measurement. If the CX emission is real, the total amount 
of oxygen in clusters might have been overestimated by up to 20\% through not accounting for charge exchange in previous
studies, skewing the star formation model of galaxy clusters. 

A final confirmation of the charge exchange sources in clusters of galaxies has to wait for a more sensitive observation with future spectroscopic instruments. The XRISM mission (launch 2021), 
equipped with the same micro-calorimeter as {\it Hitomi}, will provide the first opportunity to 
detect charge exchange in clusters. Based on the above {\it Hitomi} result, the signature Fe XXV high-$n$ transitions might be confirmed at 5$\sigma$ through a 500~ks XRISM observation of the Perseus cluster center, combined with the present {\it Hitomi} data of the same object. A same detection
confidence will be achieved by a single 100~ks Athena (launch 2030) observation. The region of
the charge exchange, as well as some physical parameters, such as the ionization degree and the 
impinging velocity of hot and cold matter, will be determined with XRISM and Athena.

\section{Progress on atomic codes}
\label{sec:5}

%$\rightarrow$ Kaastra, Gu

%5.1 review of the achievements and challenges after Hitomi (e.g., a list of representative discrepancies among codes) \\
%5.2 ongoing activities and the lab data needs 

\begin{table}[!htbp]
\caption{Fractional uncertainties on the {\it Hitomi} Perseus results}
\label{tab:syst}
\begin{threeparttable}
\centering
\begin{tabular}{cccccccc}
\hline\hline
Effect & Flux &      kT   &    Si abund.  &  Fe abund.  &    Turbulence & $N_{\rm H, hot}$   \\
\hline
statistical                  &  0    &  0   & 5\%  & 1\%  & 2\%  & 7\% \\
plasma code$^{\rm a}$        &  1\%  &  2\% & 19\% & 16\% & 10\% & 40\% \\
multi-temperature$^{\rm b}$  &  3\%  &  4\% & 11\% & 3\%  & 1\%  & 12\% \\
resonant scattering$^{\rm c}$  &  1\%  &  0   & 5\%  & 11\% & 9\%  & $-$  \\
NEI$^{\rm c}$                 &  1\%  &  1\% & 2\%  & 0    & 1\%  & 8\%  \\
charge exchange$^{\rm c}$     &   0   &  0   & 2\%  & 5\%  & 2\%  & 7\%  \\
effective area cor.$^{\rm d}$ &  1\%  &  2\% & 3\%  & 1\%  & 0    & 3\%  \\
gain cor.           &  0    &  0   & 14\% & 10\% & 3\%  & 3\%  \\
   \hline
\end{tabular}
\begin{tablenotes}
\item[$(a)$] Discrepancies on the fits using SPEX v3.03, APEC v3.0.8, and Chianti v8.0.
\item[$(b)$] Changes on the fits after applying the multi-temperature modeling of the ICM structure.
\item[$(c)$] Changes by including the resonant scattering, non-equilibrium ionization, and charge exchange
components.
\item[$(d)$] Changes from the manually corrected effective area (see \citet{2018PASJ...70...12H} for details).
\item[$(e)$] Changes from the parabolic correction on {\it Hitomi} energy scale  (see \citet{2018PASJ...70...12H} for details).
\end{tablenotes}
\end{threeparttable}
\end{table}

\begin{figure*}[!htbp]
\centering
\resizebox{0.9\hsize}{!}{\includegraphics[angle=0]{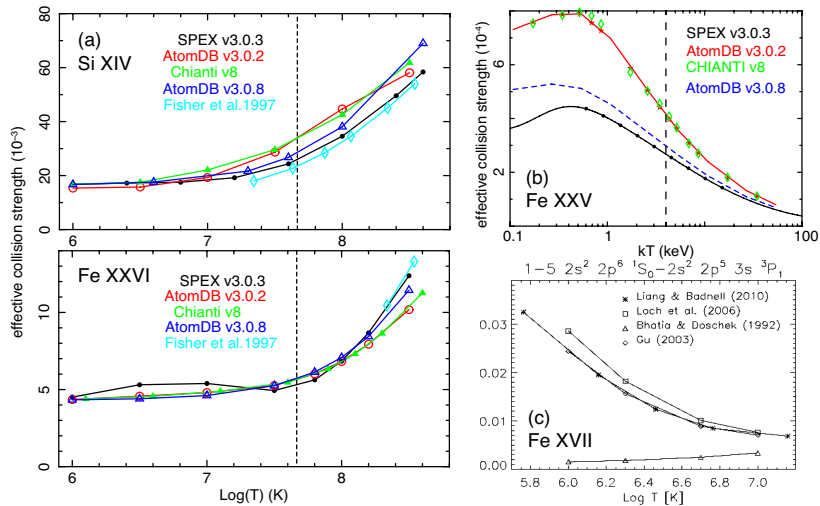}}
\caption{(a) Comparison of effective collision strength as a function of electron temperature, for
combined Ly$\alpha$1 and Ly$\alpha$2 transitions. The vertical line shows a temperature of 4~keV. (b) Same as (a), but for the 1s$^2$ ($^1$S$_0$) $-$
1s2s ($^3$S$_1$) transition of He-like Fe. (c) Effective collision 
strength for one of the major Fe XVII lines \citep{2011A&A...536A..59D}. 
\vspace{0.5cm}
}
\label{fig:cs}
\end{figure*}

The atomic data and plasma model are essential for analyzing the spectra of astronomical plasma. As mentioned earlier, the study of the 
detailed spectral features heavily relies on the theoretical calculation of atomic processes. For instance, the excitation and radiative cascade rates are crucial for the resonant scattering measurement (\S\ref{sec:1}), and the sulfur charge exchange line at $\sim 3.5$ keV is predicted based on the latest calculation of the electron-capture cross section (\S\ref{sec:4}). With the advent of high-resolution X-ray spectroscopy instruments, the need for accurate and complete atomic data and plasma model is not just growing but becoming urgent, while the production of the high quality fundamental atomic data is clearly not simple.

The {\it Hitomi}/SXS spectrum of the Perseus cluster, with $\sim 5$ eV resolution in the $2-10$ keV band, offers an unprecedented benchmark of the atomic model calculations for a $\sim 4$ keV optically-thin thermal plasma. It unveils both successes and challenges 
of the current plasma codes, the AtomDB/APEC \citep{2001ApJ...556L..91S, 2012ApJ...756..128F} and SPEX \citep{1996uxsa.conf..411K}. 
Neither the AtomDB nor SPEX could provide statistically acceptable fits to the {\it Hitomi} spectrum with their pre-launch versions \citep{2016Natur.535..117H}, immediately revealing inadequacies in atomic data of both codes. A series of updates addressed most of the issues, leading to an overall agreement between the two codes and acceptable fits to the {\it Hitomi} data with the latest versions. The remaining issues with the {\it Hitomi} data are mostly on the detailed line intensity \citep{2018PASJ...70...12H}. As shown in Fig.~\ref{fig:cs}, the calculations of the collisional excitation and dielectronic recombination for H/He/Li-like ions highly depend on the theoretical models, which do
not converge well for a 4~keV plasma. The resulting systematic uncertainties on the Fe abundance is 16\%. As shown in Table~\ref{tab:syst}, 
it is 16 times higher than the statistical uncertainty, and yet significantly higher than the instrumental calibration errors and the errors caused by astrophysical modeling. The final solution to the issue is still waiting for a targeted laboratory benchmark measurement.

The atomic code issues revealed by the {\it Hitomi} data are only the tip of the iceberg, as a large portion of the database remains untested. A long-term and continuous effort is required to ensure that plasma codes will be ready to face the challenges from 
the future high-resolution spectroscopy, which will be achieved with the XRISM and the Athena mission. 
These missions will 
observe a broad range of cosmic plasmas, including collisionally-ionized plasma with temperatures between $10^{4} - 10^{9}$ K, either
optically-thin and optically-thick, and photo-ionized plasma with a similar range of ionization states. The
source could be in either ionization equilibrium and non equilibrium. The non-thermal excitation (\S\ref{sec:2}) 
and charge exchange (\S\ref{sec:4}) line emission are also expected from various types of targets.  

Among the atomic calculations to be verified, the Fe L-shell spectrum (with Fe XVII to Fe XXIV) is one of the top priorities, 
as it contains strong 
emission lines which are crucial for a range of science cases (\S\ref{sec:1}). To determine where the discrepancies and inadequacies
exist, the atomic team need to perform a systematic testing of the present atomic models against each other, 
as well as against deep observed data in the archives.  The recent theoretical calculations with the $R$-matrix method \citep{2010A&A...518A..64L, 2007JPhB...40.2969W} and improved distorted wave method \citep{2003ApJ...582.1241G}, as well as 
the experimental data with electron beam ion traps \citep{2012Natur.492..225B, 2016IAUFM..29A.295H} will be used for solving
these issues. When disagreement
remains, new dedicated theoretical calculation and/or targeted laboratory measurements will be required.

\section{Prospects for future missions}
\label{sec:6}

In this paper we have reviewed the X-ray spectroscopic features produced by a range of physical processes in the ICM.
Most of them are faint, easily swamped by the thermal radiation from the hot gas, and their presence can only be 
corroborated by high resolution X-ray spectroscopy. The scientific potential behind these features is
enormous, making them valid targets for the future X-ray missions. When the XRISM and Athena will be in orbit,
the resonant scattering will be studied for many clusters, and combined with the direct velocity measurement via
Doppler broadening, it will provide us a powerful probe of the anisotropy and spatial scales of the ICM random motion.
The signature of particle acceleration in merging events will be assessed, for the first time, through measurements of 
the satellite lines. The timescale needed for the ICM to equilibrate behind a merging and/or accretion shock can
be determined by using the NEI modeling. The charge exchange from the cool cores will be measured, providing a new tool to
help solve the puzzle of why there are such cold clouds embedded in otherwise hot ICM. The formation of the giant cold gas
cloud might be a key to understand the cooling vs heating balance in cluster cores.

The non-dispersive {\it Hitomi} Perseus spectrum made us aware of the problems with the present plasma codes in
modeling the thermal radiation from the ICM. Before XRISM is ready to launch, these problems will be addressed 
by a combined effort of atomic code
testing, new theoretical calculations, and targeted laboratory measurements.

\begin{acknowledgements}
L.G. is supported by the RIKEN Special Postdoctoral Researcher Program. 
SRON is supported financially by NWO, the Netherlands Organization for Scientific Research.
\end{acknowledgements}


\begin{thebibliography}{}




























\bibitem[Aharonian et al.(2017)]{2017ApJ...837L..15A} Aharonian, F.~A., Akamatsu, H., Akimoto, F., et al.\ 2017, The Astrophysical Journal, 837, L15 

\bibitem[\protect\citeauthoryear{Ahoranta et al.}{2016}]{2016A&A...592A.145A} Ahoranta J., Finoguenov A., Pinto C., Sanders J., Kaastra J., de Plaa J., Fabian A., 2016, Astronomy \& Astrophysics, 592, A145

\bibitem[Akahori \& Yoshikawa(2010)]{2010PASJ...62..335A} Akahori, T., \& Yoshikawa, K.\ 2010, Publications of the Astronomical Society of Japan, 62, 335 

\bibitem[\protect\citeauthoryear{Akimoto et al.}{1999}]{1999AN....320..283A} Akimoto F., Furuzawa A., Tawara Y., Yamashita K., 1999, Astronomische Nachrichten, 320, 283

\bibitem[\protect\citeauthoryear{Akimoto et al.}{2000}]{2000AdSpR..25..603A} Akimoto F., Furuzawa A., Tawara Y., Yamashita K., 2000, Advances in Space Research, 25, 603

\bibitem[Bernitt et al.(2012)]{2012Natur.492..225B} Bernitt, S., Brown, G.~V., Rudolph, J.~K., et al.\ 2012, Nature, 492, 225 

\bibitem[\protect\citeauthoryear{B{\"o}hringer et al.}{2001}]{2001A&A...365L.181B} B{\"o}hringer H., et al., 2001, Astronomy \& Astrophysics, 365, L181

\bibitem[Boyarsky et al.(2014)]{2014PhRvL.113y1301B} Boyarsky, A., Ruchayskiy, O., Iakubovskyi, D., \& Franse, J.\ 2014, Physical Review Letters, 113, 251301 

\bibitem[Branduardi-Raymont et al.(2007)]{2007A&A...463..761B} Branduardi-Raymont, G., Bhardwaj, A., Elsner, R.~F., et al.\ 2007, Astronomy \& Astrophysics, 463, 761 

\bibitem[Bulbul et al.(2014)]{2014ApJ...789...13B} Bulbul, E., Markevitch, M., Foster, A., et al.\ 2014, The Astrophysical Journal, 789, 13 

\bibitem[\protect\citeauthoryear{Buote, Canizares, \& Fabian}{1999}]{1999MNRAS.310..483B} Buote D.~A., Canizares C.~R., Fabian A.~C., 1999, Monthly Notices of the Royal Astronomical Society, 310, 483

\bibitem[\protect\citeauthoryear{Chandrasekhar}{1950}]{Chan50} Chandrasekhar S., 1950, Radiative Transfer, Oxford, Clarendon Press, 1950

\bibitem[\protect\citeauthoryear{Churazov et al.}{2004}]{2004MNRAS.347...29C} Churazov E., Forman W., Jones C., Sunyaev R., B{\"o}hringer H., 2004, Monthly Notices of the Royal Astronomical Society, 347, 29

\bibitem[\protect\citeauthoryear{Churazov et al.}{2010}]{2010SSRv..157..193C} Churazov E., Zhuravleva I., Sazonov S., Sunyaev R., 2010, Space Science Reviews, 157, 193

\bibitem[Conselice et al.(2001)]{2001AJ....122.2281C} Conselice, C.~J., Gallagher, J.~S., III, \& Wyse, R.~F.~G.\ 2001, The Astronomical Journal, 122, 2281 

\bibitem[Cravens(1997)]{1997GeoRL..24..105C} Cravens, T.~E.\ 1997, Geophysical Research Letters, 24, 105 

\bibitem[Cumbee et al.(2014)]{2014ApJ...787L..31C} Cumbee, R.~S., Henley, D.~B., Stancil, P.~C., et al.\ 2014, The Astrophysical Journal, 787, L31 
\bibitem[Cumbee et al.(2016)]{2016MNRAS.458.3554C} Cumbee, R.~S., Liu, L., Lyons, D., et al.\ 2016, Monthly Notices of the Royal Astronomical Society, 458, 3554 

\bibitem[Cumbee et al.(2018)]{2018ApJ...852....7C} Cumbee, R.~S., Mullen, P.~D., Lyons, D., et al.\ 2018, The Astrophysical Journal, 852, 7 

\bibitem[Dennerl et al.(2006)]{2006A&A...451..709D} Dennerl, K., Lisse, C.~M., Bhardwaj, A., et al.\ 2006, Astronomy \& Astrophysics, 451, 709 

\bibitem[\protect\citeauthoryear{Dupke \& Arnaud}{2001}]{2001ApJ...548..141D} Dupke R.~A., Arnaud K.~A., 2001, The Astrophysical Journal, 548, 141

\bibitem[\protect\citeauthoryear{Ezawa et al.}{2001}]{2001PASJ...53..595E} Ezawa H., et al., 2001, Publications of the Astronomical Society of Japan, 53, 595

\bibitem[Fabian et al.(2011)]{2011MNRAS.417..172F} Fabian, A.~C., Sanders, J.~S., Williams, R.~J.~R., et al.\ 2011, Monthly Notices of the Royal Astronomical Society, 417, 172 

\bibitem[Foster et al.(2012)]{2012ApJ...756..128F} Foster, A.~R., Ji, L., Smith, R.~K., \& Brickhouse, N.~S.\ 2012, The Astrophysical Journal, 756, 128 

\bibitem[Fujita et al.(2008)]{2008PASJ...60.1133F} Fujita, Y., Hayashida, K., Nagai, M., et al.\ 2008, Publications of the Astronomical Society of Japan, 60, 1133 

\bibitem[\protect\citeauthoryear{Gabriel \& Phillips}{1979}]{1979MNRAS.189..319G}
Gabriel, A.~H. \& Phillips, K.~J.~H. 1979, Monthly Notices of the Royal Astronomical Society, 189, 319

\bibitem[\protect\citeauthoryear{Gastaldello \& Molendi}{2004}]{2004ApJ...600..670G} Gastaldello F., Molendi S., 2004, The Astrophysical Journal, 600, 670

\bibitem[\protect\citeauthoryear{Gilfanov, Sunyaev \& Churazov}{1987}]{1987SvAL...13....3G} Gilfanov M.~R., Sunyaev R.~A., Churazov E.~M., 1987, Soviet Astronomy Letters, 13, 3

\bibitem[Gu(2003)]{2003ApJ...582.1241G} Gu, M.~F.\ 2003, The Astrophysical Journal, 582, 1241 

\bibitem[Gu et al.(2012)]{2012CaJPh..90..351G} Gu, M.~F., Beiersdorfer, P., Brown, G.~V., et al.\ 2012, Canadian Journal of Physics, 90, 351 

\bibitem[Gu et al.(2015)]{2015A&A...584L..11G} Gu, L., Kaastra, J., Raassen, A.~J.~J., et al.\ 2015, Astronomy \& Astrophysics, 584, L11 

\bibitem[Gu et al.(2016)]{2016A&A...588A..52G} Gu, L., Kaastra, J., \& Raassen, A.~J.~J.\ 2016, Astronomy \& Astrophysics, 588, A52 

\bibitem[Gu et al.(2016)]{2016A&A...594A..78G} Gu, L., Mao, J., Costantini, E., \& Kaastra, J.\ 2016, Astronomy \& Astrophysics, 594, A78 

\bibitem[Gu et al.(2018)]{2018A&A...611A..26G} Gu, L., Mao, J., de Plaa, J., et al.\ 2018, Astronomy \& Astrophysics, 611, A26 

\bibitem[\protect\citeauthoryear{Hahn \& Savin}{2015}]{2015ApJ...809..178H}
Hahn, M., \& Savin, D.~W. 2015, The Astrophysical Journal, 809, 178

\bibitem[\protect\citeauthoryear{Hamilton}{1947}]{1947ApJ...106..457H} Hamilton D.~R., 1947, The Astrophysical Journal, 106, 457

\bibitem[Hell et al.(2016)]{2016IAUFM..29A.295H} Hell, N., Brown, G.~V., Wilms, J., et al.\ 2016, IAU Focus Meeting, 29, 295 

\bibitem[Hitomi Collaboration et al.(2016)]{2016Natur.535..117H} Hitomi Collaboration, Aharonian, F., Akamatsu, H., et al.\ 2016, Nature, 535, 117 

\bibitem[Hitomi Collaboration et al.(2017)]{2017Natur.551..478H} Hitomi Collaboration, Aharonian, F., Akamatsu, H., et al.\ 2017, Nature, 551, 478 

\bibitem[Hitomi Collaboration et al.(2018)]{2018PASJ...70...10H} Hitomi Collaboration, Aharonian, F., Akamatsu, H., et al.\ 2018, Publications of the Astronomical Society of Japan, 70, 10 

\bibitem[Hitomi Collaboration et al.(2018)]{2018PASJ...70...11H} Hitomi Collaboration, Aharonian, F., Akamatsu, H., et al.\ 2018, Publications of the Astronomical Society of Japan, 70, 11 

\bibitem[Hitomi Collaboration et al.(2018)]{2018PASJ...70...12H} Hitomi Collaboration, Aharonian, F., Akamatsu, H., et al.\ 2018, Publications of the Astronomical Society of Japan, 70, 12 

\bibitem[Inoue et al.(2016)]{2016PASJ...68S..23I} Inoue, S., Hayashida, K., Ueda, S., et al.\ 2016, Publications of the Astronomical Society of Japan, 68, S23 

\bibitem[Kaastra et al.(1996)]{1996uxsa.conf..411K} Kaastra, J.~S., Mewe, R., \& Nieuwenhuijzen, H.\ 1996, UV and X-ray Spectroscopy of Astrophysical and Laboratory Plasmas, 411

\bibitem[\protect\citeauthoryear{Kaastra et al.}{1999}]{1999ApJ...519L.119K} Kaastra J.~S., Lieu R., Mittaz J.~P.~D., Bleeker J.~A.~M., Mewe R., Colafrancesco S., Lockman F.~J., 1999, The Astrophysical Journal, 519, L119

\bibitem[\protect\citeauthoryear{Kaastra et al.}{2009}]{2009A&A...503..373K} Kaastra J.~S., Bykov, A.~M. \& Werner, N., 2009, Astronomy \& Astrophysics, 503, 373

\bibitem[\protect\citeauthoryear{Kahn et al.}{2003}]{2003ASPC..301...23K} Kahn S.~M., et al., 2003, Matter and Energy in Clusters of Galaxies, ASP Conference Proceedings, 301, 23

\bibitem[Katsuda et al.(2011)]{2011ApJ...730...24K} Katsuda, S., Tsunemi, H., Mori, K., et al.\ 2011, The Astrophysical Journal, 730, 24 

\bibitem[Konami et al.(2011)]{2011PASJ...63S.913K} Konami, S., Matsushita, K., Tsuru, T.~G., Gandhi, P., \& Tamagawa, T.\ 2011, Publications of the Astronomical Society of Japan, 63, S913 

\bibitem[Lallement(2004)]{2004A&A...422..391L} Lallement, R.\ 2004, Astronomy \& Astrophysics, 422, 391 

\bibitem[Lallement(2009)]{2009SSRv..143..427L} Lallement, R.\ 2009, Space Science Reviews, 143, 427 

\bibitem[Liang \& Badnell(2010)]{2010A&A...518A..64L} Liang, G.~Y., \& Badnell, N.~R.\ 2010, Astronomy and Astrophysics, 518, A64 

\bibitem[Lisse et al.(1996)]{1996Sci...274..205L} Lisse, C.~M., Dennerl, K., Englhauser, J., et al.\ 1996, Science, 274, 205 

\bibitem[Liu et al.(2011)]{2011MNRAS.415L..64L} Liu, J., Mao, S., \& Wang, Q.~D.\ 2011, Monthly Notices of the Royal Astronomical Society, 415, L64 

\bibitem[\protect\citeauthoryear{Mathews, Buote, \& Brighenti}{2001}]{2001ApJ...550L..31M} Mathews W.~G., Buote D.~A., Brighenti F., 2001, The Astrophysical Journal, 550, L31

\bibitem[Mitchell et al.(1976)]{1976MNRAS.175P..29M} Mitchell, R.~J., Culhane, J.~L., Davison, P.~J.~N., \& Ives, J.~C.\ 1976, Monthly Notices of the Royal Astronomical Society, 175, 29P 

\bibitem[\protect\citeauthoryear{Molendi et al.}{1998}]{1998ApJ...499..608M} Molendi S., Matt G., Antonelli L.~A., Fiore F., Fusco-Femiano R., Kaastra J., Maccarone C., Perola C., 1998, The Astrophysical Journal, 499, 608

\bibitem[Mushotzky et al.(1996)]{1996ApJ...466..686M} Mushotzky, R., Loewenstein, M., Arnaud, K.~A., et al.\ 1996, The Astrophysical Journal, 466, 686 

\bibitem[\protect\citeauthoryear{Ogorzalek et al.}{2017}]{2017MNRAS.472.1659O} Ogorzalek A., et al., 2017, Monthly Notices of the Royal Astronomical Society, 472, 1659

\bibitem[\protect\citeauthoryear{Panagoulia, Sanders, \& Fabian}{2015}]{2015MNRAS.447..417P} Panagoulia E.~K., Sanders J.~S., Fabian A.~C., 2015, Monthly Notices of the Royal Astronomical Society, 447, 417

\bibitem[Peterson et al.(2001)]{2001A&A...365L.104P} Peterson, J.~R., Paerels, F.~B.~S., Kaastra, J.~S., et al.\ 2001, Astronomy and Astrophysics, 365, L104 

\bibitem[Pinto et al.(2016)]{2016MNRAS.461.2077P} Pinto, C., Fabian, A.~C., Ogorzalek, A., et al.\ 2016, Monthly Notices of the Royal Astronomical Society, 461, 2077 

\bibitem[\protect\citeauthoryear{de Plaa et al.}{2012}]{2012A&A...539A..34D} de Plaa J., Zhuravleva I., Werner N., Kaastra J.~S., Churazov E., Smith R.~K., Raassen A.~J.~J., Grange Y.~G., 2012, Astronomy and Astrophysics, 539, A34

\bibitem[\protect\citeauthoryear{Rephaeli et al.}{2008}]{2008SSRv..134...71R}
Rephaeli, Y., Nevalainen, J., Ohashi, T. \& Bykov, A.~M., 2008, Space Science Reviews, 134, 71

\bibitem[Roberts \& Wang(2015)]{2015MNRAS.449.1340R} Roberts, S.~R., \& Wang, Q.~D.\ 2015, Monthly Notices of the Royal Astronomical Society, 449, 1340 

\bibitem[Russell et al.(2012)]{2012MNRAS.423..236R} Russell, H.~R., McNamara, B.~R., Sanders, J.~S., et al.\ 2012, Monthly Notices of the Royal Astronomical Society, 423, 236 

\bibitem[\protect\citeauthoryear{Sakelliou et al.}{2002}]{2002A&A...391..903S} Sakelliou I., et al., 2002, Astronomy and Astrophysics, 391, 903

\bibitem[\protect\citeauthoryear{Sanders \& Fabian}{2006}]{2006MNRAS.370...63S} Sanders J.~S., Fabian A.~C., 2006, Monthly Notices of the Royal Astronomical Society, 370, 63

\bibitem[\protect\citeauthoryear{Sazonov, Churazov \& Sunyaev}{2002}]{2002MNRAS.333..191S} Sazonov S.~Y., Churazov E.~M., Sunyaev R.~A., 2002, Monthly Notices of the Royal Astronomical Society, 333, 191

\bibitem[Shah et al.(2016)]{2016ApJ...833...52S} Shah, C., Dobrodey, S., Bernitt, S., et al.\ 2016, The Astrophysical Journal, 833, 52 

\bibitem[\protect\citeauthoryear{Shang \& Oh}{2013}]{2013MNRAS.433.1172S} Shang C., Oh S.~P., 2013, Monthly Notices of the Royal Astronomical Society, 433, 1172

\bibitem[Smith et al.(2001)]{2001ApJ...556L..91S} Smith, R.~K., Brickhouse, N.~S., Liedahl, D.~A., \& Raymond, J.~C.\ 2001, The Astrophysical Journal Letters, 556, L91 

\bibitem[Smith et al.(2012)]{2012AN....333..301S} Smith, R.~K., Foster, A.~R., \& Brickhouse, N.~S.\ 2012, Astronomische Nachrichten, 333, 301 

\bibitem[Snowden et al.(2004)]{2004ApJ...610.1182S} Snowden, S.~L., Collier, M.~R., \& Kuntz, K.~D.\ 2004, The Astrophysical Journal, 610, 1182 

\bibitem[Stancil(2001)]{2001ASPC..247....3S} Stancil, P.~C.\ 2001, Spectroscopic Challenges of Photoionized Plasmas, 247, 3 

\bibitem[Takahashi et al.(2016)]{2016SPIE.9905E..0UT} Takahashi, T., Kokubun, M., Mitsuda, K., et al.\ 2016, Proceedings of the SPIE, 9905, 99050U 

\bibitem[Walker et al.(2015)]{2015MNRAS.453.2480W} Walker, S.~A., Kosec, P., Fabian, A.~C., \& Sanders, J.~S.\ 2015, Monthly Notices of the Royal Astronomical Society, 453, 2480 

\bibitem[\protect\citeauthoryear{Werner et al.}{2009}]{2009MNRAS.398...23W} Werner N., Zhuravleva I., Churazov E., Simionescu A., Allen S.~W., Forman W., Jones C., Kaastra J.~S., 2009, Monthly Notices of the Royal Astronomical Society, 398, 23

\bibitem[Witthoeft et al.(2007)]{2007JPhB...40.2969W} Witthoeft, M.~C., Whiteford, A.~D., \& Badnell, N.~R.\ 2007, Journal of Physics B Atomic Molecular Physics, 40, 2969 

\bibitem[Wong et al.(2011)]{2011ApJ...727..126W} Wong, K.-W., Sarazin, C.~L., \& Ji, L.\ 2011, The Astrophysical Journal, 727, 126 

\bibitem[\protect\citeauthoryear{Xu et al.}{2002}]{2002ApJ...579..600X} Xu H., et al., 2002, The Astrophysical Journal, 579, 600

\bibitem[Del Zanna(2011)]{2011A&A...536A..59D} Del Zanna, G.\ 2011, Astronomy and Astrophysics, 536, A59 

\bibitem[Zhang et al.(2014)]{2014ApJ...794...61Z} Zhang, S., Wang, Q.~D., Ji, L., et al.\ 2014, The Astrophysical Journal, 794, 61 

\bibitem[\protect\citeauthoryear{Zhuravleva et al.}{2010}]{2010MNRAS.403..129Z} Zhuravleva I.~V., Churazov E.~M., Sazonov S.~Y., Sunyaev R.~A., Forman W., Dolag K., 2010, Monthly Notices of the Royal Astronomical Society, 403, 129

\bibitem[\protect\citeauthoryear{Zhuravleva et al.}{2011}]{2011AstL...37..141Z} Zhuravleva I.~V., Churazov E.~M., Sazonov S.~Y., Sunyaev R.~A., Dolag K., 2011, Astronomy Letters, 37, 141

\bibitem[\protect\citeauthoryear{Zhuravleva et al.}{2013}]{2013MNRAS.435.3111Z} Zhuravleva I., et al., 2013, Monthly Notices of the Royal Astronomical Society, 435, 3111



\end{thebibliography}


\end{document}